# Properties of Neutral Charmed Mesons in Proton–Nucleus Interactions at 70 GeV

(SVD-2 Collaboration)

A.N. Aleev, E.N. Ardashev, A.G. Afonin, V.P. Balandin, S.G. Basiladze, S.F. Berezhnev, G.A. Bogdanova, M.Yu. Bogolyubsky, A.M. Vishnevskaya, V.Yu. Volkov, A.P. Vorobiev, A.G. Voronin, G.G. Ermakov, P.F. Ermolov[†], S.N. Golovnia, S.A. Gorokhov, V.F. Golovkin, N.I. Grishin, Ya.V. Grishkevich, V.N. Zapolsky, E.G. Zverev, S.A. Zotkin, D.S. Zotkin, D.E. Karmanov, V.I. Kireev, A.A. Kiriakov, V.N. Kramarenko, A.V. Kubarovsky, N.A. Kouzmine, L.L. Kurchaninov, G.I. Lanshikov, A.K. Leflat, S.I. Lyutov, M.M. Merkin, G.Ya. Mitrofanov, V.S. Petrov, Yu .P. Petukhov, A.V. Pleskach, V.V. Popov, V.M. Ronjin, V. N. Ryadovikov[*], D.V. Savrina, V.A. Senko, M.M. Soldatov, L.A. Tikhonova, N.F. Furmanec, A.G. Kholodenko, Yu.P. Tsyupa, N.A. Shalanda, A.I. Yukaev, V.I. Yakimchuk.

**Abstract** — The results of treatment of data obtained in the SERP-E-184 experiment "Investigation of mechanisms of the production of charmed particles in proton–nucleus (pA) interactions at 70 GeV and their decays" by irradiating, with a beam of 70-GeV protons, the active target of the SVD-2 facility, this target consisting of carbon, silicon, and lead plates, are presented. After separating a signal from the two-particle decay of neutral charmed mesons and estimating the cross section for charm production at a threshold energy $\sigma(c\bar{c}) = 7.1 \pm 2.4(\text{stat.}) \pm 1.4(\text{syst.})$ μb/nucleon, some properties of $D^0$ and $\check{D}^0$ are investigated. These include the dependence of the cross section on the target mass number (its A dependence); the behavior of the differential cross sections $d\sigma/dp_t^2$ and $d\sigma/dx_F$; and the dependence of the parameter $\alpha$ on the kinematical variables $x_F$, $p_t^2$, and $p_{\text{lab}}$. The experimental results in question are compared with predictions obtained on the basis of the FRITIOF7.02 code.

## INTRODUCTION

The SERP-E-184 experiment "Investigation of mechanisms of the production of charmed particles in proton–nucleus (pA) interactions at 70 GeV and their decays" [1] is being performed at the SVD- 2 facility of the Institute for High Energy Physics (IHEP, Protvino). An active target, which consists of carbon, silicon, and lead plates, is exposed to a beam of 70-GeV protons. Upon separating a signal in the effective-mass spectrum of the $K\pi$ system, an estimate of the cross section for charmed-meson production in pA interactions at energies in the threshold region was given in [2]. The cross section for charm production proved to be

$$\sigma(c\bar{c}) = 7.1 \pm 2.4(\text{stat.}) \pm 1.4(\text{syst.}) \text{ μb/nucleon.}$$

The value obtained for the cross section exceeds the predictions of hard QCD [1] ($\sigma(c\bar{c}) \sim 1$ μb). At the same time, variations in the parameters of the model within possible limits change the error field [3], with the result that this cross section does not seem overly large (Fig. 1a was taken from [4], and our point was added). Attempts at estimating the cross section for charm production at energies in the threshold region were undertaken more than 20 years ago at the IHEP facility BIS-2 in irradiating a carbon target with neutrons of energy in the range 40–70 GeV [5]. In the kinematical region $x_F>0.5$, the measured cross section for $D^0$-meson production proved to be considerably larger than theoretical predictions, namely, $\sigma(D^0)=28\pm14$ μb/nucleus. Upon rescaling it to the entire kinematical region the cross section for charm production proved to be about 5 μb/nucleon. Approximately the same theoretical estimate of this quantity was obtained by Kaidalov's group in calculating the charm-production cross section on the basis of the model of quark–gluon strings (QGS model or QGSM) [6]. Figure 1b shows the graph taken from [6] and supplemented with our point.

---

[†] deceased
[*] riadovikov@ihep.ru

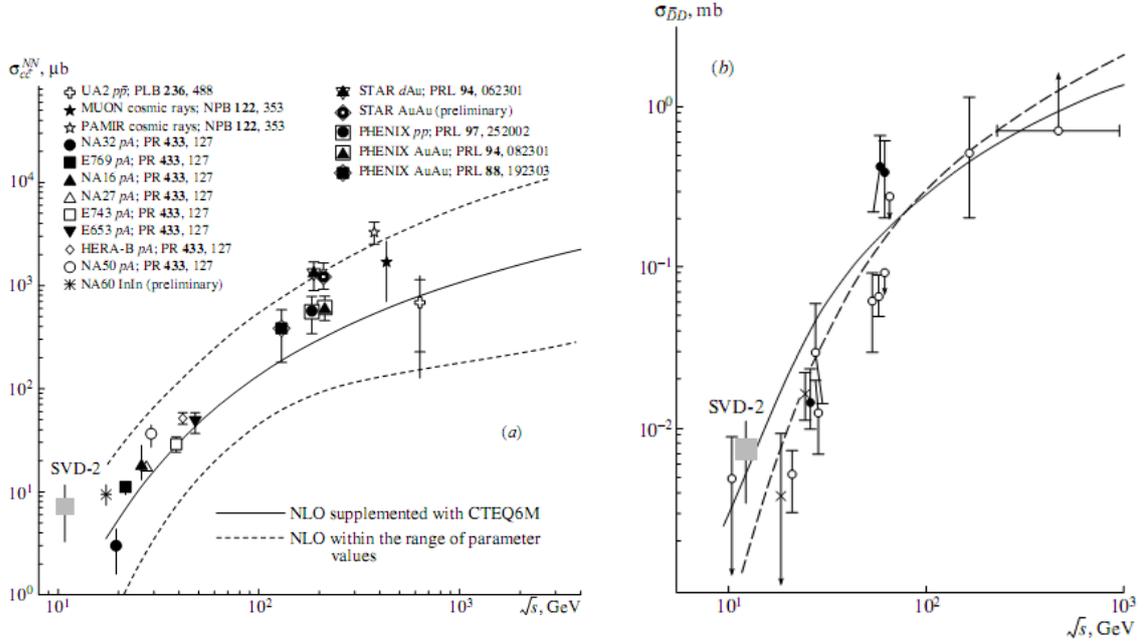

Fig. 1. Experimental cross sections for charm production in pA interactions along with theoretical predictions based on (a) perturbative QCD [3] or (b) the model of quark–gluon strings [6].

A detailed description of the SVD-2 facility can be found in [1]. The presence in the E-184 experiment of a target containing carbon, silicon, and lead plates makes it possible to measure the dependence of the charm-production cross section on the atomic weight (A) of target nuclei. In [2], it was shown that, in this experiment, the parameter α in the A dependence ($\sigma \sim A^\alpha$) is $1.08 \pm 0.12$, which is consistent with results of other experiments [7–9].

A detailed simulation of processes accompanying the detection of charmed-particle decays at the SVD-2 facility on the basis of the FRITIOF7.02 and GEANT3.21 packages makes it possible to determine the efficiencies of all procedures of the data treatment system and their dependence on the kinematical parameters $p_t^2$ and $x_F$; this in turn permits deducing estimates for inclusive spectra of neutral D mesons.

## LIFETIME OF NEUTRAL D MESONS

In order to verify whether the separated $K\pi$ decays are the decays of charmed mesons, we measured their lifetime on the basis of the dependence of the cross section for the reaction $pA \rightarrow D^0 + X$ on the range of the $K\pi$ system. The visible range was corrected by multiplying it by the factor (p/M); that is, use was made of $L=L_{vis}/(p/M)$, where p is the momentum of the system and M is its measured mass. The interval of ranges was broken down into subintervals (see Fig. 2); in each subinterval, the effective-mass spectrum was constructed for the $K\pi$ system, and the cross section was determined from the number of events in the signal from $D^0$-meson decay. Because of the smallness of the statistical data sample, signals from $D^0$ and $\check{D}^0$ mesons were united into a single spectrum. In approximating the dependence of the cross section on the range (see Fig. 2) by a function of the form $\sigma \sim \exp(-L/c\tau)$, one obtains the fitted value of $c\tau = 0.123 \pm 0.024$ mm, which agrees with the tabulated value of 0.124 mm within the errors.

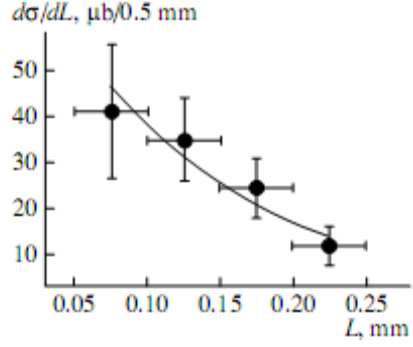

Fig. 2. Cross section for the production of neutral D mesons as a function of their range: (points) experimental data and (curve) result of parametrizing data by an exponential function.

## DIFFERENTIAL CROSS SECTION $d\sigma/dp_t^2$

The acceptance of the SVD-2 facility makes it possible to measure the transverse momentum ($p_t$) and the Feynman variable ($x_F = 2p_L/\sqrt{s}$) of charmed mesons over a broad region spanned by $p_t^2$ between 0 and 4 (GeV/c)$^2$ and $x_F$ between –0.2 and 0.6. A simulation shows that 54% of $D^0$ mesons and 23% of $\check{D}^0$ mesons then fall within the spectrometer aperture.

In order to obtain $p_t^2$ spectra, we constructed effective-mass spectra of the $K\pi$ system in four $p_t^2$ intervals. In each spectrum, we determined the number of events, $N_{det}$, involving the decay of neutral D mesons and calculated the inclusive partial cross section for the respective $p_t^2$ interval by the formula

$$\sigma(D^0)_{nucl} = K_{instr} \cdot N_{det} \cdot A^{0.7}/(Br \cdot \varepsilon \cdot L_{int}),$$

employing values determined previously for the efficiencies and other quantities (branching ratio, integrated luminosity, and instrumental factor) [2]. For four $p_t^2$ intervals, Table 1 gives values of respective cross sections along with statistical errors in them. In order to calculate the cross-section value averaged over nuclei, we used the summed signal and the averaged atomic weight of the nuclei in just the same way as was described in [2]. The measured average value of the transverse momentum of neutral D mesons is $<p_t>=1.02$ GeV/c. Approximating the transverse-momentum dependence of the experimental cross section for all nuclei by the expression $d\sigma/dp_t^2 \sim \exp(-bp_t^2)$, we obtained the fitted exponent value of $b = 0.79 \pm 0.15$ (GeV/c)$^{-2}$ (Fig. 3).

Table 1. Cross sections for $D^0$-meson production for four $p_t^2$ intervals ($\Delta p_t^2 = 1.0$ (GeV/c)$^2$).

| $<p_t^2>$, (GeV/c)$^2$ | $\varepsilon_{det}$, % | $d\sigma$, μb/nucleus | | | |
|---|---|---|---|---|---|
| | | carbon | silicon | lead | average over nuclei |
| 0.5 | 3.7 | 13 ± 13 | 83 ± 28 | 945 ± 285 | 218 ± 45 |
| 1.5 | 3.8 | 26 ± 18 | 63 ± 24 | 669 ± 237 | 157 ± 38 |
| 2.5 | 3.4 | 15 ± 15 | 30 ± 17 | 281 ± 162 | 72 ± 27 |
| 3.5 | 3.5 | 14 ± 14 | 10 ± 10 | 91 ± 91 | 20 ± 14 |

In pA collisions, we investigated the behavior of the A-dependence parameter α as a function of kinematical variables. Despite small statistics of the signal and large errors associated with this, we made an attempt at observing the $p_t^2$ dependence of α. For four $p_t^2$ intervals, Fig.4a shows the differential cross section as a function of the target atomic weight. One can see that the slopes of the respective straight lines are different for different values of $p_t^2$. The experimental data in question are indicative of a decrease in the parameter α with increasing $p_t^2$ according to an exponential law (see Fig. 4b).

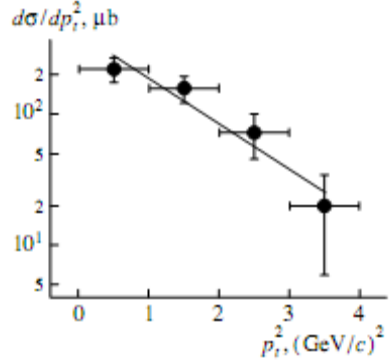

Fig. 3. Differential cross section $d\sigma/dp_t^2$ for the production of neutral D mesons: (points) experimental data and (curve) exponential function fitted to the data.

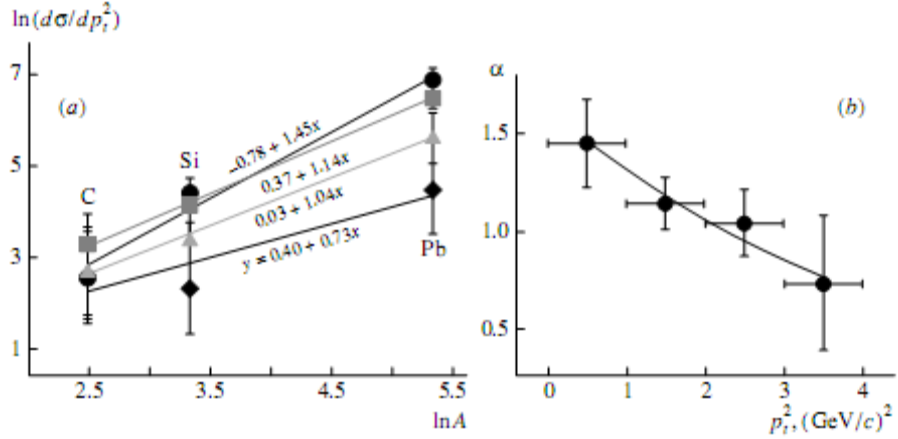

Fig. 4. (a) Differential cross section as a function of the atomic weight of target nuclei and (b) parameter α as a function of $p_t^2$: (points) experimental data and (curve) exponential function fitted to the data.

## DIFFERENTIAL CROSS SECTION $d\sigma/dx_F$

We have studied the behavior of the cross section for the reaction $pA \rightarrow D^0 + X$ as a function of the Feynman variable $x_F$. The cross sections for different $x_F$ intervals were calculated by a method similar to that used in studying the transverse-momentum dependence - that is, in constructing the effective-mass spectra of the $K\pi$ system in four $x_F$ intervals and in determining the number of events in the signal for each interval (see Table 2).

Figure 5 shows experimental values of the cross sections for the production of neutral charmed mesons versus the variable $x_F$. In order to describe the respective dependence, we employed a standard parametrization of the form $d\sigma/dx_F \sim (1-|x_F|)^n$. The fitted value of the parameter n proved to be 6.8±0.8, and the respective mean value was $<x_F> = 0.12$.

Following the same line of reasoning as in studying the $p_t^2$ dependence of the parameter α, we have analyzed the dependence of this parameter on the variable $x_F$. For this, we estimated signals from $D^0$ mesons and the corresponding cross sections for their production in $x_F$ intervals for three materials of the active target (see Fig. 6a).

Table 2. Cross section for $D^0$-meson production for various $x_F$ intervals ($\Delta x_F = 0.2$)

| $\langle x_F \rangle$ | $\varepsilon_{det}$, % | $d\sigma$, µb/nucleus | | | |
|---|---|---|---|---|---|
| | | carbon | silicon | lead | average over nuclei |
| −0.1 | 2.6 | 10 ± 10 | 13 ± 13 | 245 ± 173 | 40 ± 23 |
| 0.1 | 9.4 | 16 ± 9 | 55 ± 14 | 541 ± 135 | 123 ± 21 |
| 0.3 | 13.5 | 7 ± 5 | 15 ± 6 | 118 ± 52 | 39 ± 10 |
| 0.5 | 12.5 | 2 ± 2 | 6 ± 4 | 25 ± 25 | 6 ± 4 |

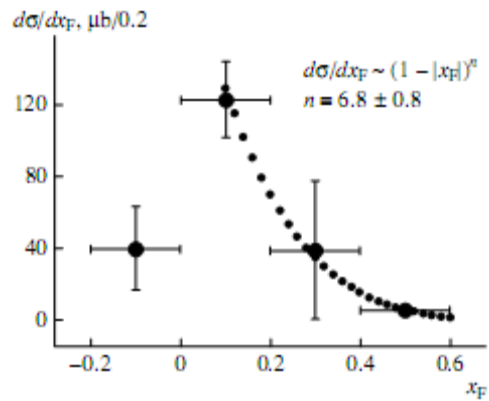

Fig. 5. Differential cross section for the production of neutral D mesons, $d\sigma/dx_F$: (points) experimental data and (dotted curve) fitted parametrization of the form $d\sigma/dx_F \sim (1-|x_F|)^n$.

From Fig. 6b, one can see that the parameter α decreases with increasing $x_F$. If we describe the data in terms of an exponential function, then, for $x_F \to 1$, the parameter α decreases to the value of 0.55. This agrees with the theoretical prediction made in [6].

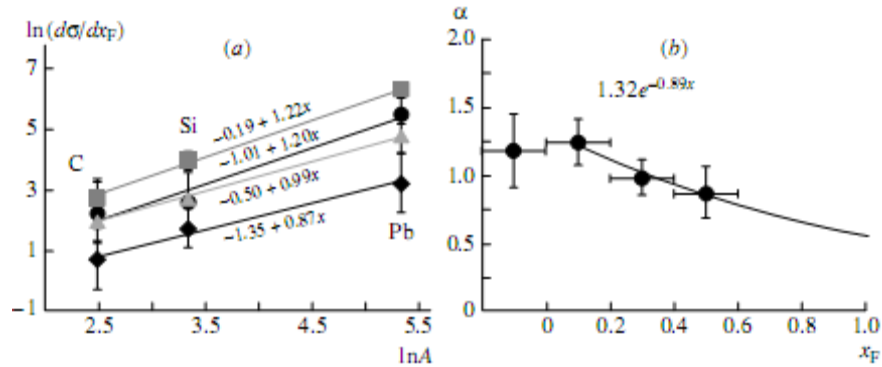

Fig. 6. (a) Differential cross section $d\sigma/dx_F$ as a function of the target atomic weight for four intervals of $x_F$ (see Table 2) and (b) parameter α as a function of $x_F$. The displayed points stand for experimental data, while the curve represents the fitted parametrization by an exponential function.

It is noteworthy that we estimate here systematic uncertainties in values obtained for the cross sections in question at a level of 20% of the statistical errors in them.

# FRITIOF CODE AND A DEPENDENCE OF CROSS SECTIONS

The Lund string model is used in the FRITIOF code for simulating hadron–hadron and hadron–nucleus interactions. It is assumed that, after 4-momentum exchange, hadrons become two excited string states, which, thereupon, emit gluons in the approximation of QCD color dipoles. The ultimate hadronization is performed by using the Lund string-fragmentation model. A collision with a nucleus is treated as independent collisions of an incident nucleon with constituent nucleons of the nucleus involved. The Fermi motion of nucleons, the deformation of the nucleus, and multiple rescattering are taken into account. The nucleon-distribution density in a nucleus is described by the Woods–Saxon potential. We employed this code to perform a model investigation of the dependence of the parameter α on the kinematical parameters of $D^0$ mesons and to compare the results obtained in this way with experimental data. For three values of the target atomic weight (C, Si, and Pb), the available numbers of simulated (MC) events involving $D^0$ mesons were weighed in such a way that the A dependence with parameter α =1 was satisfied on average for all events. From three distributions with respect to a given kinematical variable ($x_F$, $p_t^2$, and $p_{lab}$), the dependence of the parameter α on this quantity was calculated thereupon for $D^0$ mesons.

According to [10], the $x_F$ dependence of α must reflect the contribution to the cross section from various nuclear subprocesses, such as final-state absorption, interaction with closely flying hadrons (interactions with comovers), the shadowing of parton distributions, the parton energy loss in the medium, and the effect of intrinsic-charm components. This leads to an increase or a decrease in the parameter α as xF increases.

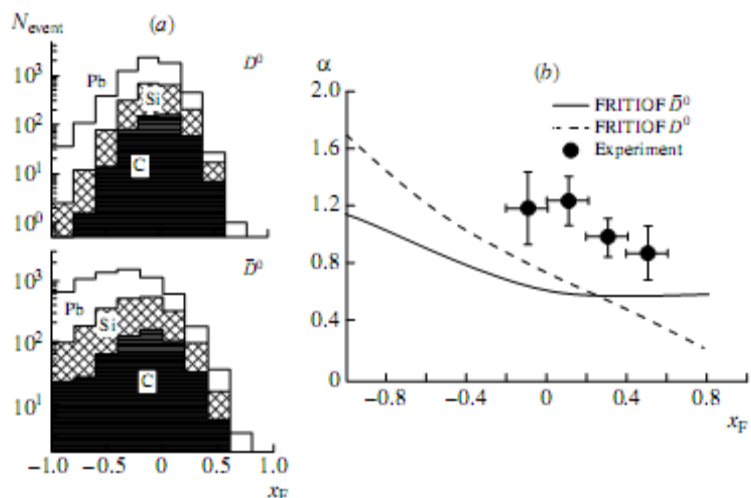

Fig. 7. (a) Distributions of $x_F$ for D mesons and (b) parameter α as a function of $x_F$. The displayed points stand for experimental data, and the curves represent the results of the simulation.

Figure 7a shows the input simulated distributions of events in three targets (C, Si, Pb) with respect to the Feynman variable $x_F$ for $D^0$ and $\check{D}^0$ mesons. On the basis of these distributions, we calculate the parameter α. The graphs representing its dependence on the variable $x_F$ are shown in Fig. 7b (solid and dashed curves). Also given in that figure are the experimental values of α for four $x_F$ intervals.

It is worth noting that the variable $x_F$ may take "unphysical" values for MC events - those going beyond the range [–1, 1]. This is because, in calculating the Feynman variable $x_F=2p_L/\sqrt{s}$, the c.m. energy $\sqrt{s}$ proves to be underestimated if the interaction of the incident nucleon with several nucleons of the target nucleus is not taken into account. In [11], it was shown that, upon taking into account all interacting nucleons of the nucleus (in the case of the simulation on the basis of the FRITIOF code, this number is known), the distribution of the variable $x_F$ falls within the range [–1, 1], as it must. Unfortunately, the number of interacting nucleons of the nucleus is unknown in the experiment; therefore, the c.m. energy is calculated for two nucleons (incident and target), and one has to employ unphysical values of the variable $x_F$ for MC events. Concurrently, α decreases with increasing $x_F$ over the entire range of $x_F$, and the experiment confirms this qualitatively.

Figure 8 shows $p_t^2$ distributions for simulated $D^0$ and $\check{D}^0$ mesons (Fig. 8a) and the $p_t^2$ dependence of the parameter α (Fig. 8b). A comparison of FRITIOF-simulated dependences and experimental points reveals that there is no even qualitative agreement between the model and the experiment (experimental errors are large because of small statistics).

In Fig. 9, we present similar distributions with respect to $p_{lab}$ of neutral D mesons. In this case, there is no problem in displaying data, in contrast to what we have for the Feynman variable $x_F$, in which case the number of interacting nucleons of the nucleus is unknown. (The same dependence of the parameter α is presented in [12]). Here, we see qualitative agreement between the experiment and the model; that is, α decreases as $p_{lab}$ of neutral D mesons increases.

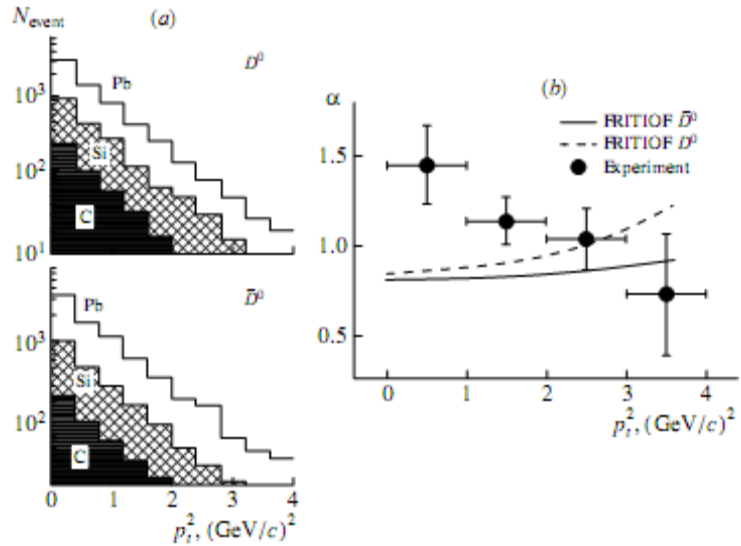

Fig. 8. (a) Distributions of $p_t^2$ for D mesons and (b) parameter α as a function of $p_t^2$. The displayed points stand for experimental data, and the curves represent the results of the simulation.

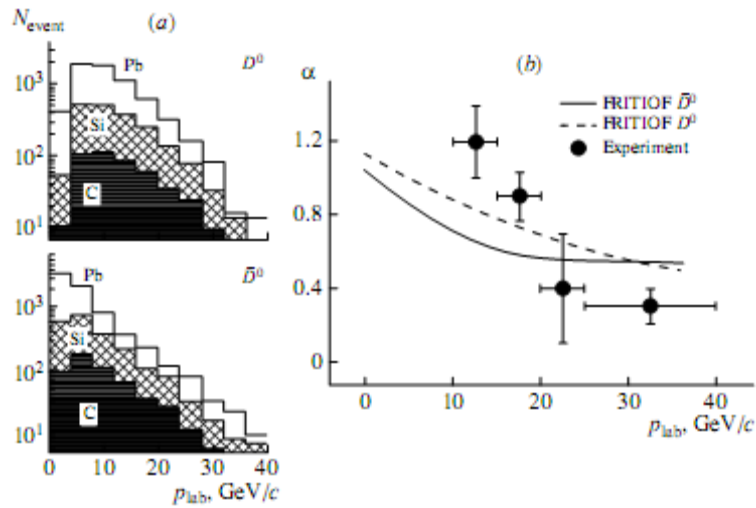

Fig. 9. (a) Distributions of $p_{lab}$ for D mesons and (b) parameter α as a function of $p_{lab}$. The displayed points stand for experimental data, and the curves represent the results of the simulation.

## CONCLUSIONS

In conclusion, we present Table 3, which gives the results of some experiments devoted to studying charm production in pA interactions. One can see that, within the errors, our results are compatible with those data. However, further investigations aimed at refining the properties of charmed particles produced in pA interactions at energies in the threshold region are required.

In comparing the behavior of the A-dependence parameter α of the cross section as a function of kinematical variables in FRITIOF-simulated events and experimental data, we can see qualitative agreement for the case of the Feynman variable $x_F$ of neutral D mesons and their $p_{lab}$. In the case of the variable $p_t^2$, there is a significant discrepancy: within the model, α is virtually independent of $p_t^2$, but the experimental points are indicative of a decrease in α with increasing $p_t^2$.

Table 3. Data on the production of neutral D mesons and on their properties in pA interactions.

| Experiment | Beam energy, GeV | $\sigma(D^0)$, $\mu$b/nucleon | $\alpha$ ($\sigma \sim A^\alpha$) | $n$ ($d\sigma/dx_F \sim (1-x_F)^n$) | $b$ ($d\sigma/dp_t^2 \sim \exp(-bp_t^2)$) |
|---|---|---|---|---|---|
| **SVD-2** | 70 | 7.1 ± 3.8 | 1.08 ± 0.12 | 6.8 ± 0.8 | 0.79 ± 0.15 |
| E769 [7] | 250 | 12.0 ± 3.8 | 0.92 ± 0.08 | 4.1 ± 0.6 | 0.95 ± 0.09 |
| NA16 [7] | 360 | 20.4 ± 16.0 | – | – | – |
| NA27 [7] | 400 | 18.3 ± 2.5 | – | 4.9 ± 0.5 | 1.0 ± 0.1 |
| E-789 [8] | 800 | 17.7 ± 4.2 | 1.02 ± 0.05 | – | 0.91 ± 0.12 |
| E743 [7] | 800 | 22.0 ± 14.0 | – | 8.6 ± 2.0 | 0.8 ± 0.2 |
| E653 [7] | 800 | 39.0 ± 15.0 | – | 11.0 ± 2.0 | 1.1 ± 0.2 |
| HERA-B [9] | 920 | 48.7 ± 10.6 | 0.97 ± 0.07 | 7.5 ± 3.2 | 0.84 ± 0.1 |

## ACKNOWLEDGMENTS

This work was supported by the Russian Foundation for Basic Research (project no. 09-02-00445) and was funded by a grant (no. 1456-2008-2) for support of leading scientific schools.